**Theoretical prediction of Curie temperature in two-dimensional ferromagnetic monolayer**

Yufei Xue[1], Zhong Shen[1], Zebin Wu[1] and Changsheng Song[1,2†].

[1]Key Laboratory of Optical Field Manipulation of Zhejiang Province, Department of Physics, Zhejiang Sci-Tech University, Hangzhou 310018, China
[2]Longgang Institute of Zhejiang Sci-Tech University, Wenzhou 325802, China.
†Email: cssong@zstu.edu.cn

**ABSTRACT**

Theoretical prediction of Curie temperature ($T_C$) is of vital importance for designing the spintronic devices in two-dimensional (2D) ferromagnetic materials. Herein, based on the extensive investigation of Monte Carlo simulations, we summary and propose an improved method to estimate $T_C$ more precisely, which includes the different contributions of multiple near-neighbor interactions. Taking monolayer $CrI_3$ as an example, the trends of $T_C$ with biaxial strain are investigated via Monte Carlo simulations, mean-field formulas and our method. Besides, our method is not only accurate and convenient to predicting the $T_C$ in 2D ferromagnetic honeycomb lattice $CrI_3$ but it can be extended for predicting the $T_C$ of other 2D lattices. Our work paves the way to accelerate the prediction and discovery of novel 2D ferromagnets for spintronic applications.

## I. INTRODUCTION

The Curie temperature ($T_C$) is one of the essential properties of two-dimensional (2D) ferromagnetic (FM) materials [1-4]. Moreover, the method of predicting $T_C$ is extremely important not only for the deep understanding of the phase transition from ferromagnetic to paramagnetic phase but also for further exploring novel 2D FM materials in spintronic devices[5-7]. Typically, mean-field and Heisenberg exchange theories are commonly used to explain the process of ferromagnetic phase transformation. These two theories do lead to different pictures of certain aspects of the collective characteristics and the ferromagnetic phase transformation. Mean-field theory considers the complex interactions in the magnet as a powerful "molecular field" (Weiss field), which exists inside the magnet, and that an individual particle should respond to the average field generated by interactions with surrounding particles [8]. While Heisenberg exchange theory considers long-range exchange interactions between atomic spins mediated by the overlap of the atomic orbitals in the absence of an applied field.

Many efforts have been done to develop an accurate theoretical method for calculating $T_C$. However, most methods based on mean-field theory, generally overestimate the $T_C$ of 2D ferromagnets due to their failure to consider the sufficient number of the contributions of near-neighbor magnetic interactions [9-13]. Monte Carlo (MC) simulation is recognized as a more accurate method for systems where stochastic processes occur [14,15], but the results of MC are largely affected by the initial lattice size and the number of MC steps per site, leading to a large amount of time cost [16]. Thus, a reliable method for theoretically predicting $T_C$ of 2D ferromagnets remains a serious challenge. Early studies showed that the magnitude of $T_C$ is strongly influenced by the exchange interactions, lattice shapes, and spins. For a determined lattice type of a crystal with spin S, its $T_C$ is proportional to exchange interaction $J$ [17]. Additionally, for the simple body-centered and face-centered cubic lattices, the formula $K_B T_C = 5(Z-1)(11X-1)/96J$ reproduces the estimated $T_C$ accurately, where $X = S(S+1)$ and Z is the lattice coordination-number [18]. However, the neglect of relatively complex correlation among spin moments leads to a large deviation between the theoretical magnetic behavior and the experiments near $T_C$. Therefore, there is the potential significance for developing an accurate method to predict $T_C$ under an appropriate assumption.

According to the established Mermin-Wagner theorem [19], long-range ferromagnetism is not almost existent in strictly 2D isotropic systems. Excitingly, long-range FM order in the 2D limit has been recently experimentally confirmed in monolayer $CrI_3$ ($T_C \approx 45$ K). Meanwhile, theoretical results have been reported that the $T_C$ of monolayer $CrI_3$ is 107 K by Ising model [20,21], and 109 K by mean-field theory [22,23], indicating that Ising model and mean-field theory may overestimate $T_C$ a lot [24]. Thus, based on abundant MC studies, we propose a dependable method (3NN-MF) for theoretical prediction of $T_C$ with considering first-, second-, and third-nearest neighboring exchange interactions, which relies on the parameterization of first-principles calculations to construct a 2D ferromagnetic Heisenberg Hamiltonian.

Our paper is structured as follows. Firstly, based on the results of MC simulations, we propose a more efficient method to estimate $T_C$ that incorporates multiple near-neighbor exchange interactions and the single-ion anisotropy (SIA), which is obtained from DFT calculations. To eliminate potential finite-size effects of $T_C$, the periodic boundary conditions and the Binder cumulant are adopted in our calculation. Furthermore, by taking monolayer $CrI_3$ as an example, we investigate the physical origin of observed magnetic behavior under biaxial strain, and then compare our results with two mean-field methods and MC simulations. These studies will provide a guideline for strain-induced modulation of $T_C$ in 2D ferromagnets and prediction of the high $T_C$ materials ahead of experiments.

## II. CALCULATIONAL DETAILS

The density-functional theory (DFT) calculations are performed using projected augmented wave (PAW) method as implemented in the Vienna ab initio Simulation Package (VASP) [25,26]. In all of these calculations, we use the generalized gradient exchange-correlation function of the Perdew-Burke-Ernzerhof (PBE) [27] flavor. The Brillouin zone is sampled by 6 × 6 × 1 k-point grid mesh [28] and a plane wave cutoff energy of 400 eV is used. In addition, a vacuum of 15 Å is applied along the z-axis to avoid any artificial interactions between images. Relaxations are performed until the Hellmann-Feynman force on each atom becomes smaller than 0.05 eV/Å and the total energy is converged to be within $10^{-5}$ eV. Spin-polarization is considered to reproduce the semiconducting nature of this system. For the biaxial strain study of $CrI_3$, the lattice constants are changed accordingly for each strain while the atomic positions are fully optimized. In addition, the exchange interactions and the SIA for strained cells are obtained from DFT calculations. Schematizations of crystal structures are prepared by using VESTA [29].

Moreover, the temperature evolutions of the magnetization and susceptibility are obtained by Monte Carlo simulations using the Hamiltonian of Eq. (1). In order to eliminate potential finite-size effects in calculating the $T_C$, we apply the periodic boundary conditions in our calculation, and then the Binder cumulant [30] of $CrI_3$ with different lattice sizes also calculated for a finite size scaling analysis. In addition, to obtain reliable simulated data with high accuracy, we start with ferromagnetic configurations and use 600000 sweeps to sufficiently thermalize the system into equilibrium at each temperature, and all statistical results are obtained from next 80000 sweeps.

According to spin Heisenberg model including the magnetic anisotropy factor, we can obtain the magnetic parameters of 2D ferromagnets. The Hamiltonian equation can be written as

$$H_{spin} = \sum_{i,j} J_n S_i S_j + \sum_i A(S_i^z)^2, \tag{1}$$

where $S_{i(j)}$ is the spin operator on site i (j), $J_n$ is the *n*th near-neighbor exchange interaction between sites i and j, $S_i^z$ is the spin component parallel to the z-direction, and *A* is the easy-axis single-ion anisotropy (SIA). The different near neighbors exchange interactions $J_{ij}$ and *A* can be calculated based on DFT with local spin density approximation plus Hubbard parameter U level. In addition, we find that it is reasonable to take U = 1.0 eV when investigate the magnetic properties of $CrI_3$ with biaxial strain (as shown in Figs. S1 and S2 of supplementary material).

## III. RESULTS AND DISCUSSION

### A. Curie temperature in 2D ferromagnet

Previously, the Mermin-Wagner theorem states that long-range ferromagnetic ordering cannot exist in a 2D isotropic Heisenberg spin system with a finite temperature [19]. However, in most 2D FM materials, the atomic arrangement and energy of different directions are anisotropic. The SIA is several orders of magnitude weaker than the exchange interaction in most 2D FM systems, meaning that the Curie temperature is not sensitive to it in light of energy[31]. In a relatively small range of $A/J_1$: 0.004 ~ 0.5, the error of $K_B T_C / (J_1 |S|^2)$ is less than 0.4, as shown in Fig. 1(a). Nonetheless, we should emphasize that the magnetic anisotropy does play an indispensable role in a long-range magnetic stability under thermal condition.

To make a better prediction of $T_C$ in 2D materials, based on the theoretical foundation of mean-filed theory and MC study, we propose an empiric method called 3NN-MF for the prediction of $T_C$. Firstly, the classical Heisenberg spin Hamiltonian of FM material is given by:

$$E_{FM} = \sum_{n=1}^{\infty} \left[ Z_n J_n S_i S_j + A(S_i^z)^2 \right] + E_0, \qquad (2)$$

where $Z_n$ is the lattice coordination-number and $E_0$ is the non-magnetic atomic energy (not sensitive for different magnetic states). When $J_n$ is negative, its magnitude is proportional to the $T_C$ [32].

Similarly, the corresponding antiferromagnetic (AFM) energy can be written as:

$$E_{AFM} = \sum_{n=1}^{\infty} \left[ (Z_n - 2m) J_n S_i S_j + A(S_i^z)^2 \right] + E_0, \qquad (3)$$

where *m* is the number of the neighboring anti-parallel spin, which is related to the spin at site i. According to Eqs. (2) and (3), the energy difference of phase transition from FM to AFM state can be written as:

$$E_{FM} - E_{AFM} = 2 \sum_{n=1}^{\infty} m J_n |S|^2. \qquad (4)$$

Additionally, we then explore the correlation between the multiple exchange interactions $J_n$ and $T_C$. It is well known that $T_C$ is proportional to the $J_1$ [32,33], which is also consistent with the results of Fig. 1(b). The relationship between $J_1$, $J_2$, $J_3$ and $T_C$ are still well nearly linear dependent. Thus, it is possible to find that the magnetic contributions of multiple nearest neighbor interactions to obtain the approximated formula:

$$K_B T_C \approx -\sum_{n=1}^{\infty} \beta_n J_n |S|^2, \qquad (5)$$

where $\beta_n$ are a set of coefficients and we can obtain the values of $\beta_n$ for a certain type of lattice by setting the supercell sizes (L×L) between L = 4 and L =112. If we examine the relationship between L and $\beta_1$, from Fig. 1(c), we will find that $\beta_1$ (L→∞) is smaller than that of in a finite size. Moreover, one can extrapolate the values of $\beta_1$ with the thermodynamic limit (L→∞). Furthermore, as shown in Fig. 1(d), we can obtain $\beta_1 \approx$ 0.40, 0.63 and 1.20, $\beta_2 \approx$ 1.52, 1.00, 1.42 and $\beta_3 \approx$ 0.60, 1.67, 1.87 for 2D honeycomb lattice, simple cubic lattice, and hexagonal close-packed lattice, respectively. For example, in 2D hexagonal close-packed of $MnBi_2Te_4$, we adopt the parameters $J_1 = -0.92$ meV, S = 1, and $Z_1 = 6$, substitute them into Eq. (5) and then estimate the $T_C$ is 11.85 K, which is in good agreement with the MC result of $T_C \approx$ 12.0 K [34].

Alternatively, the $T_C$ also can be roughly estimated according to Eqs. (4) and (5) as the following (it is convenient for preliminary estimate but not for great accuracy):

$$K_B T_C \approx \frac{-\beta_1}{2m}(E_{FM} - E_{AFM}). \qquad (6)$$

The $T_C$ is estimated by the energy difference between $E_{FM}$ and $E_{AFM}$ for 2D FM materials. Obviously, it is different from the traditional mean-filed formulas $T_C=2(E_{DLM} - E_{FM})/(3K_B)$ [35,36] or $T_C=2(E_0 - E_{FM})/(3K_B)$ [16], which estimate $T_C$ by considering the energy difference between disordered local moment (DLM) $E_{DLM}$ and ferromagnetic state $E_{FM}$, and the exchange interactions $J$ can be obtained from $E_{DLM} - E_{FM}$ or $E_0 - E_{FM}$ in many previous works [16,35,36]. It should be noticed, at the Curie point, the thermal fluctuations are required to balance with the energy difference between FM and DLM states, there is still some local magnetization, the traditional mean-field formulas does not capture the local magnetic order above the transition temperature, leading to the overestimation of $T_C$ [37]. Therefore, it is evident that the neglect of local magnetic order may give rise to lager deviations of theoretical predicting $T_C$ [38-41].

On the other hand, from a lattice structure point of view, the ferromagnetism may vanish ($E_{AFM} - E_{FM} \approx 0$) at a certain strain (the inevitable result of the competition between direct interactions and superexchange interactions), which means that the $T_C$ is close to zero at the critical transition point between FM and AFM, and this similar conclusion has also been widely reported in other material [42-44]. Thus, we can obtain the $T_C$ by the energy difference between FM and various AFM configurations: $K_B T_C \propto E_{AFM} - E_{FM}$.

## B. Curie temperature in $CrI_3$

Next, we will focus on theoretical prediction of $T_C$ and explore the effects of biaxial strain on the magnetic properties in monolayer CrI$_3$. Figure. 2(a) shows the crystal structure of monolayer CrI$_3$. The Cr ions form a honeycomb network sandwiched by two atomic planes of I atoms. As shown in Fig. 2(b), four different magnetic configurations are considered to evaluate the magnetic ground state by comparing their total energies, when the spin orientations are all set to +z direction to represent FM configuration, while the adjacent Cr ion spins in the z-axis are anti-parallel to represent three different AFM configurations: Néel-AFM, Stripy-AFM and Zigzag-AFM, respectively.

Here, we take 2D honeycomb magnetic lattice as an example to predict $T_C$. According to Eqs. (2) and (3), a 1×2×1 supercell contains four magnetic atoms as shown in Fig. 2, the corresponding energy difference per unit cell can be derive as:

$$E_{FM} - E_{AFM}^{Néel} = (12J_1 + 0J_2 + 12J_3)|S|^2/4,$$
$$E_{FM} - E_{AFM}^{Stripy} = (8J_1 + 16J_2 + 0J_3)|S|^2/4,$$
$$E_{FM} - E_{AFM}^{Zigzag} = (4J_1 + 16J_2 + 12J_3)|S|^2/4. \quad (7)$$

As mentioned above, the Curie temperature can be preliminarily determined by Eq. (6). When $n = 1$, for Néel-AFM, Stripy-AFM and Zigzag-AFM configurations, $m = 3, 2, 1$, respectively, $\beta_1 = 0.40$ for $Z_1 = 3$. Generally speaking, $E_{FM} - E_{AFM}^{Zigzag}$ tends to achieve the minimum energy difference between FM and AFM configurations when $T_C$ is close to 0 at certain strain, and the exchange interactions ($J_2$ and $J_3$) with a longer distance are much weaker than the first-nearest neighbor interactions ($J_1$). Therefore, we can estimate the $T_C$ by the energy difference between $E_{FM}$ and $E_{AFM}^{Zigzag}$: $K_B T_C \approx -0.2(E_{FM} - E_{AFM}^{Zigzag})$. For more accurate studies, $J_2$ and $J_3$ should be considered in 3NN-MF formula. With S = 3/2 in monolayer CrI$_3$, we can get the following formula by Eq. (5) and the results of Fig. 1(c):

$$K_B T_C \approx -0.90J_1 - 3.42J_2 - 1.35J_3. \quad (8)$$

we notice that $\beta_2/\beta_1 \approx 3.8$ (L→∞) in honeycomb lattice, $\beta_2$ is just about four times as large as $\beta_1$, which is in good agreement with $K_B T_C \propto E_{AFM}^{Zigzag} - E_{FM}$ (the coefficient of $J_2$ is four times higher than that of $J_1$ in Eq. (7).

As a starting point, we investigate the $T_C$ dependence of biaxial strain. The strain ε is defined as $\varepsilon = (a-a_0)/a_0 \times 100\%$, where $a_0$ and a are the lattice constants for the unstrained and strained systems, respectively. As shown in Fig. 3 (a), the optimized lattice constant $a_0$ of monolayer CrI$_3$ is 7.00 Å. Both the first-nearest neighboring Cr-Cr bond length and Cr-I-Cr bond angle are increasing approximately linearly as the growing biaxial strain. According to the well-known Goodenough-Kanamori-Anderson rules [45-47], the super-exchange interaction among Cr-I-Cr is the direct consequence of the FM state, especially for materials with the bond angles of super-exchange close to 90°. The bond angle of CrI$_3$ with ε = 0 is around 98°, which probably induces a ferromagnetic super-exchange interaction and makes it FM order.

Then, we investigate the energy difference $\Delta E$ ($\Delta E = E_{FM} - E_{AFM}$) to study the magnetic ground state dependence of strain in CrI$_3$. As illustrated in Fig. 3(b), the FM state of CrI$_3$ favors stable from $\varepsilon = -3\%$ to $+6\%$. By calculating the energies of FM state and three AFM states, one also can determine the exchange interactions $J_1$, $J_2$, and $J_3$ with Eq. (7). As shown in Fig. 4, the magnitude of $|J_1|$ and $|J_2|$ decrease with increasing strain, while the third-nearest neighboring exchange interaction ($J_3 > 0$) maintains AFM coupling, which weaken the ferromagnetism as strain increases. As shown in Table 1, the magnetic ground states are Néel-AFM with strain larger than $\sim -3\%$, which are also consistent with the previous studies [48,30]. Notably, the drastic variations of $J_1$ (and $J_3$) for AFM states are probably attributable to the occurrence of phase transitions between FM and AFM states. These similar changes have been reported in monolayer CrGaTe$_3$ [49].

As we known, in most case, the lattice size of 2D ferromagnets is easily altered by the biaxial strain, but the lattice symmetry (the numbers of the $n$th-nearest neighbors of magnetic atoms) hardly changes. Thus, neglecting the effect of lattice slight deformation which induced by the biaxial strain, we can estimate the $T_C$ of CrI$_3$ via our proposed 3NN-MF method. To further investigate the dependence of biaxial strain on $T_C$, and verify the validity of 3NN-MF method [Eq. (8)], here three different methods to predict $T_C$ of CrI$_3$ are chosen for comparison in Fig. 5: MC simulations [50], mean-field theory (MFT) formula: $K_B T_C = Z_1 J_1 S(S+1)/3$ and mean-field approximation (MFA) formula: $K_B T_C = 3/2\, J_1$ [32,33]. The strain-free $T_C$ of CrI$_3$ calculated by our 3NN-MF method is 43.89 K, and the values calculated by MFT and MFA are 92.29 K and 36.91 K, respectively. Obviously, our method is more closer to that of the experimental result of $T_C = 45$ K [51,52]. Notably, we derive $K_B T_C = 15/4\, J_1$ based on MFT ($Z_1$=3, S=3/2 for CrI$_3$), which apparently overestimates the $T_C$ a lot because of the neglect of the disordered local moments. However, if we consider a simple average $<\cos^2\theta> = 1/2$, where $\theta$ is the angle between spin and z direction, the MFT would be $K_B T_C = 15/8\, J_1$, and gives an acceptable $T_C \approx 46.14$ K. However, the MFT and MFA have their considerable limitations where they only include the effect of nearest exchange interaction $J_1$. As we discussed earlier, $T_C$ is linearly dependent for $J_n$: $T_C \propto -J_1 - \alpha J_2$, we obtain the coefficients $\alpha$ of $J_2$ is 3.42, whereas the $\alpha$ are 3/2 for MFA and 15/4 for MFT, respectively. Particularly, $\beta_2$ is about four times larger than $\beta_1$, suggesting that the contribution of $J_2$ cannot be ignored for a reliable prediction. Meaningfully, our work is significantly important for searching high-temperature 2D ferromagnets.

The SIA is essential for the stabilization of magnetic order in 2D materials. Since there are two magnetic atoms in a primary cell of CrI$_3$, the SIA can be estimated as follows: SIA = MAE / (2|S|$^2$), where the MAE is the magnetic anisotropy energy (MAE), which is defined as the difference between energies corresponding to the magnetization in in-plane and off-plane directions (MAE = $E_\parallel - E_\perp$). The SIA of CrI$_3$ is $\sim -0.369$ meV (as shown in Fig. S1 of supplemental material), which is consistent with the previous results [23,53,54]. The $T_C$ is $\sim 47.87$

K for CrI$_3$ via Binder cumulant crossing method [30] obtained from the crossing point of the different curves that proper averaging with the large system size L of 50, 60, and 70 (as shown in Figs. S3 (a) and S3 (b) of supplemental material). Moreover, the $T_C$ is nearly unchanged when L is greater than 50.

However, we should emphasize that the 3NN-MF formula and the mean field formulas adopted in our manuscript have the limitations that the neglect of magnetic anisotropy and dipole-dipole interaction could cause the deviation of calculated $T_C$. Generally speaking, magnetic anisotropy and dipole-dipole interaction play merely the role of a "trigger" in the Curie temperature, which is not sensitive to them according to our MC study and previous studies [31]. The 3NN-MF formula can give an approximate $T_C$ with a small magnetic anisotropy. For ferromagnets with a certain larger magnetic anisotropy, the MFT and MFA formula may also give a reasonable value of $T_C$ with an accepting error. Besides, we only focus on the Curie temperature of 2D ferromagnets monolayer, do not investigate it for antiferromagnets, because our assumption [Eq. (5)] is only suitable for ferromagnet monolayer and it is not consistent with the mean-filed formula and MC results for antiferromagnet. Moreover, the exchange interactions $J_n$ are related to the energy differences between the FM and AFM configurations, but this processing may not be applicable to itinerant magnetic metals which fit the Stoner model rather than the Heisenberg model to understand the mechanism yielding the magnetic order [55].

## IV. SUMMARY

In summary, based on MC simulations with anisotropic Heisenberg model in 2D ferromagnets, a theoretical study is conducted on the relationship between multiple exchange interactions and $T_C$, and we investigate the $\beta_n$ for different lattice types by fitting the MC results to predict $T_C$. As an example, we calculate $T_C$ of monolayer CrI$_3$ under a range of biaxial strain by MC simulations, mean-field formulas and our 3NN-MF formula. The exchange interactions $J_1$, $J_2$, $J_3$ and SIA for different strains obtained from DFT calculations, especially $J_2$ is important and should be included in the $T_C$ estimation. Under the assumptions we proposed, the 3NN-MF formula including the magnetic contributions of multiple nearest-neighbor interactions are of reliable and convenient. In addition, the method we proposed can be extended to the $T_C$ estimation for other types of 2D ferromagnetic lattices. Our work is expected to theoretical predict $T_C$ and to give insight into future research to improve the $T_C$ of intrinsic 2D ferromagnetic materials.

## SUPPLEMENTARY MATERIAL

See supplementary material for more information of the effect of Hubbard U on single-ion anisotropy, exchange interactions, as well as Curie temperature and the results for the Binder cumulants in monolayer CrI$_3$.

## DATA AVAILABILITY

The data used to support the findings of this study are available within the article and its supplementary material.

## ACKNOWLEDGMENTS

This work was supported by National Natural Science Foundation of China (No.11804301), the Natural Science Foundation of Zhejiang Province (No.LY21A040008), and the Fundamental Research Funds of Zhejiang Sci-Tech University (No.2021Q043-Y, LGYJY2021015).

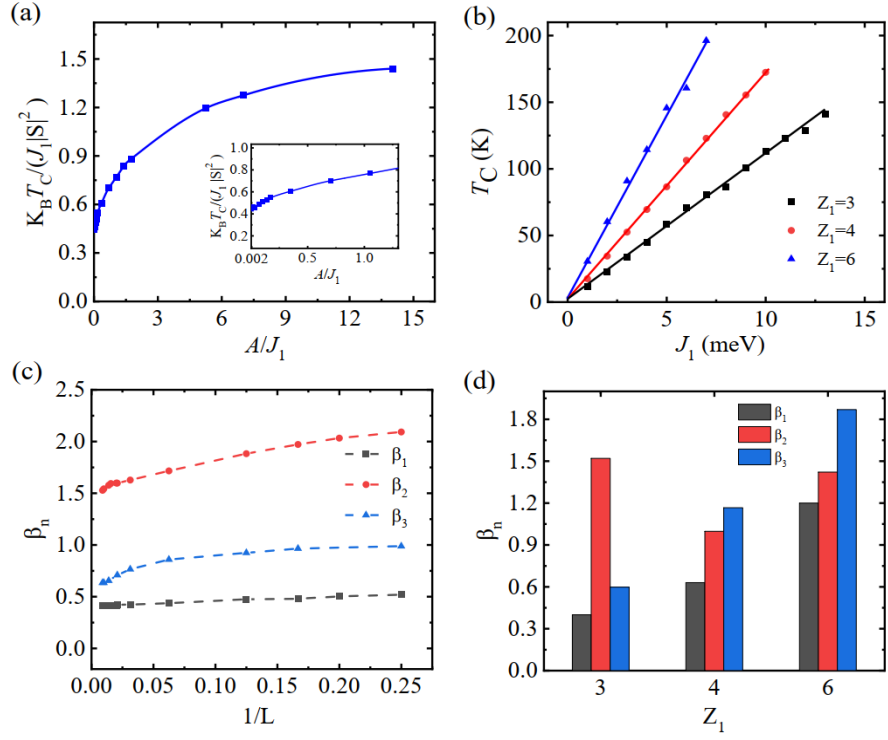

**FIG. 1.** The Curie temperature as functions of (a) single-ion anisotropy A and (b) the first-nearest neighboring exchange interactions $J_1$ in different 2D lattices with S = 3/2. (c) The constants $\beta_1$, $\beta_2$ and $\beta_3$ as a function of 1/L for honeycomb lattice, where $\beta_n = K_B T_C / (J_n |S|^2)$ derived from Eq. (5). (d) The variation of $\beta_1$, $\beta_2$ and $\beta_3$ with 2D honeycomb lattice ($Z_1 = 3$), simple cubic lattice ($Z_1 = 4$), and hexagonal close-packed lattice ($Z_1 = 6$) are represented by black, red and blue bars, respectively.

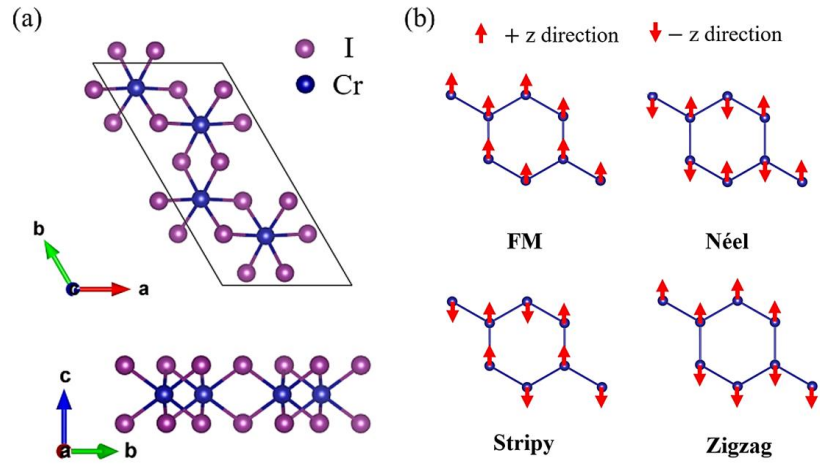

**FIG. 2.** (a) Top and side views of monolayer $CrI_3$. The black solid lines indicate the supercell of 1×2×1. (b) The four magnetic orders, namely FM, Néel-AFM, Stripy-AFM, Zigzag-AFM, and non-magnetic atoms are not shown in schematic diagram.

**TABLE 1.** The exchange interactions($J_1$, $J_2$ and $J_3$), magnetic ground states, and transition temperature ($T_{C/N}$ for Curie/Néel temperatures) in $CrI_3$. The last 2 columns are the MC and 3NN-MF results.

| Strain (%) | $J_1$ (meV) | $J_2$ (meV) | $J_3$ (meV) | Ground state | $T_{C/N}$(MC)(K) | $T_C$(3NN-MF)(K) |
|---|---|---|---|---|---|---|
| −4   | 17.23 | −0.61 | −5.98 | Néel | 121.43 | - |
| −3.9 | 16.78 | −0.61 | −5.83 | Néel | 118.01 | - |
| −3.5 | 15.25 | −0.59 | −5.36 | Néel | 107.14 | - |
| −3   | −2.66 | −0.60 | 0.19  | FM   | 53.22  | 50.91 |
| −2   | −2.64 | −0.57 | 0.16  | FM   | 52.32  | 50.40 |
| 0    | −2.12 | −0.52 | 0.11  | FM   | 47.87  | 43.89 |
| 2    | −1.46 | −0.48 | 0.10  | FM   | 34.97  | 34.53 |
| 4    | −1.04 | −0.43 | 0.09  | FM   | 27.36  | 28.31 |
| 6    | −0.64 | −0.40 | 0.06  | FM   | 23.38  | 23.03 |

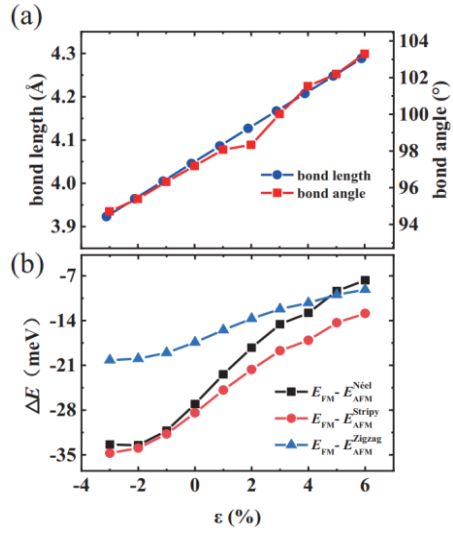

**FIG. 3.** (a) The first-nearest neighboring Cr-Cr bond length and Cr-I-Cr bond angle, (b) energy difference $\Delta E$ (meV) between FM ($E_{FM}$) and AFM ($E_{AFM}^{Néel}$, $E_{AFM}^{Stripy}$ and $E_{AFM}^{Zigzag}$) configurations of CrI$_3$ in dependence of biaxial strain.

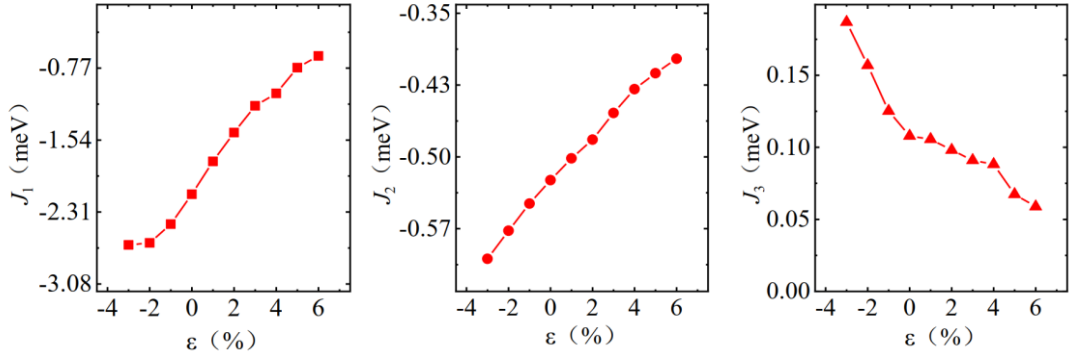

**FIG. 4.** The evolutions of the three exchange interactions $J_1$, $J_2$ and $J_3$ as a function of biaxial strain in monolayer $CrI_3$.

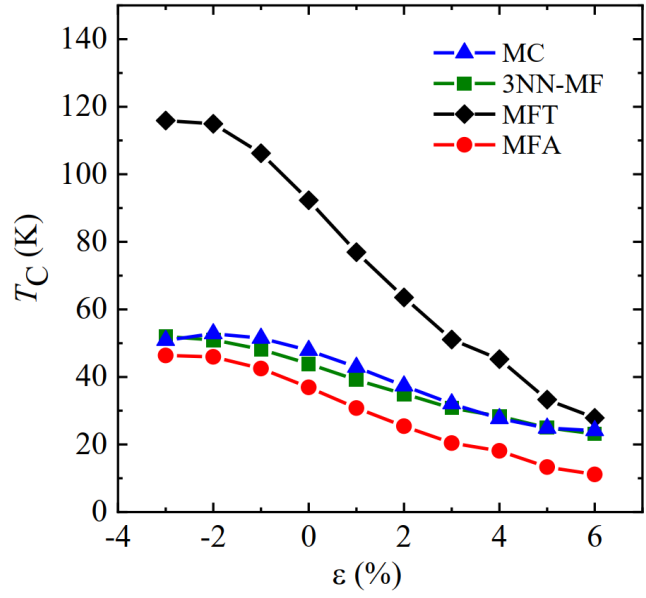

**FIG. 5.** Curie temperatures of CrI$_3$ as a function of biaxial strain by means of MC simulations, 3NN-MF method, and two mean-field approximation formulas (MFT, MFA).